\documentclass[aps,prl,letterpaper,11pt,twoside,tightenlines,nofootinbib,showpacs,preprint,twocolumn]{revtex4}
\usepackage{graphicx}
\usepackage[sort&compress]{natbib}
\usepackage{latexsym}
\usepackage{epsfig}
\begin{document}
\newcommand{\eg}{{\it e.g.}}
\newcommand{\etal}{{\it et. al.}}
\newcommand{\ie}{{\it i.e.}}
\newcommand{\be}{\begin{equation}}
\newcommand{\dd}{\displaystyle}
\newcommand{\ee}{\end{equation}}
\newcommand{\bea}{\begin{eqnarray}}
\newcommand{\eea}{\end{eqnarray}}
\newcommand{\bef}{\begin{figure}}
\newcommand{\eef}{\end{figure}}
\newcommand{\bce}{\begin{center}}
\newcommand{\ece}{\end{center}}
\def\lsim{\mathrel{\rlap{\lower4pt\hbox{\hskip1pt$\sim$}}
    \raise1pt\hbox{$<$}}}         
\def\gsim{\mathrel{\rlap{\lower4pt\hbox{\hskip1pt$\sim$}}
    \raise1pt\hbox{$>$}}}         

\title{Inhomogeneous phase of a Gluon Plasma  at finite temperature and 
 density}
\author{P.~Castorina and D.Zappal\`a}
\affiliation{Dipartimento di Fisica, Universit\`a di Catania,
 and INFN Sezione di Catania, Via Santa Sofia 64
I-95123 Catania, Italia}

\date{\today}
\begin{abstract}
By considering the non perturbative effects associated with the fundamental modular region, a
new phase of a Gluon Plasma at finite  density is proposed. It corresponds to the 
transition from glueballs to  
non perturbative gluons  which condense at a non vanishing momentum. In this respect  the proposed phase 
is analogous to the color superconducting LOFF phase for fermionic systems. 
 
\end{abstract}
 \pacs{25.75.-q,  25.75.Dw,  25.75.Nq}
 \maketitle

Quantum Chromodynamics (QCD) under extreme conditions has been intensively 
studied  and a rich phase diagram in $T - \mu$ plane is now well established.

At  small chemical potential, there is a critical temperature, $T_c$,  where
the string tension goes to zero and there is  a crossover to the deconfined quark - gluon plasma (QGP).
Moreover the lattice simulations of the QGP phase transition  clearly indicate non perturbative effects above $T_c$
\cite{boyd1,boyd2,kac,aoki1,aoki2,karsch1,asakawa1,asakawa2,digiacomo1,digiacomo2}, that is:
a) the Stefan-Boltzman limits for the pressure and the energy density of the system are  not yet 
reached  at $T \simeq 4 T_c$; b) correlated $\bar q q$ bosonic  pairs survive 
up to $T \simeq 2.3 T_c$; c) for $T<T_c$ the gluon condensate is temperature independent
whereas  for $T>T_c$ its chromoelectric part rapidly decreases and, up to $2 T_c$,  
the chromomagnetic contribution remains almost constant.
On the other side of the phase diagram, i.e. at small temperature and 
above a critical quark chemical potential, $\mu_c$, 
there is a transition to a color  superconducting phase,  which turns out
to be  stable in the so called LOFF phase, 
where the fermion pairs condense in a state with total momentum $\vec q \ne 0$
\cite{now,rev,roberto1,raja1,raja2,ciminale,fuku,max1}. 

Phenomenological quasi-particle models have been applied to fit lattice results
 \cite{levai,pescier,paolo1,paolo2} and, in particular,
for a gluon plasma (GP), i.e. a
pure SU(3) gauge theory at finite temperature, 
lattice data  on  pressure and energy density can be 
fitted  \cite{zwanziger1}, for $T > 2 T_c$,
by a gluon gas  with the following gluon dispersion relation
\be
\label{eq1}
E(k) = \sqrt{\vec k^2 + \frac{M^4}{\vec k^2}}
\ee
where $M \simeq 0.7 $ Gev \cite{zwanziger1}.
Equation (\ref{eq1}) has been derived in \cite{gribov,zwanziger2} after an analysis
of the physical configurations of a non abelian gauge theory, once the Coulomb gauge condition,
$\partial_i A_i=0$, is fixed. This condition still leaves the freedom of having 
gauge equivalent configurations (or Gribov copies) which should not be counted 
when enumerating the physical states. The corresponding reduction to the physical states only,
the so called Fundamental Modular Region (FMR), has the effect of changing the massless dispersion
relation of the gluon, $E(k)=|\vec k|$, into Eq.(\ref{eq1}).

The dispersion relation in  Eq.(\ref{eq1}) is extremely interesting not only for the theoretical reasons 
associated with confinement \cite{zwanzigern} but also because  the energy has a minimum at a finite value of the
momentum, $|\vec k|_g = M $, and increases   both in the infrared and in the ultraviolet regions of the momentum.
These are exactly the general conditions discussed by Brazovskii \cite{brazovsky} which imply  a stripe phase,
i.e. a boson condensate in a state with $\vec k \ne 0$. 
This approach has been formulated for phase transitions in condensed matter \cite{condensed}, which 
have  been experimentally tested \cite{verifica}.

Moreover,  infrared and ultraviolet behaviors analogous to  Eq.(\ref{eq1})
are typical in non-commutative self-interacting scalar field theory \cite{gubser, noi, bietenolz} 
with the consequence 
that  spontaneous symmetry breaking cannot occur 
for a homogeneous background but only for  an inhomogeneous phase, where bosons condense at
non zero momentum, analogously to the LOFF phase for fermion pairs.

Therefore, a non vanishing minimum gluon energy,  as predicted from Eq.(\ref{eq1}),
opens the possibility of studying non trivial dynamical effects also at finite  density. 
In particular, in the analysis of a pure SU(3) gauge theory at finite temperature and density, 
the natural question arises if the dispersion relation in Eq.(\ref{eq1})
implies that at  large density there is a gluon stripe phase. 
In this letter we shall  address  this issue.

Since from  Eq.(\ref{eq1}) the minimum energy 
of the gluon  is $E^*_g = \sqrt{2} M$, 
then at $T=0$ and $\mu=0$  we can derive 
a rough estimate of glueball mass,
where two valence gluons are bound by a one gluon effective exchange interaction \cite{ken}. 
Thus the corresponding glueball mass is 
\be
\label{eq2}
M_G \simeq 2 E^*_g - \frac{\alpha_s(r)}{r}
\ee
where the typical scale of the bound state is $r \simeq 1/M$ and  the one loop expression for $\alpha_s(r)$
has been used, with $\Lambda_{QCD} \simeq 200$ Mev.
For the fitted value of $M$ in \cite{zwanziger1}, $M = 0.7$ Gev, $M_g \simeq 1.5$ Gev. 
This suggests that the
zero momentum glueballs condensate, associated with $<\alpha_s/\pi G_{\mu \nu}^a G^{a \mu \nu}>_0 \ne 0$,
corresponds to confined gluons of  energy
$E^*_g$ or minimum momentum  $|\vec k|_g = M $ and that,
by increasing  the density, one can expect ( see for example ref. \cite{alessandro})
a transition from a glueballs condensate to a deconfined, but still non perturbative, gluonic phase 
with condensation in the mode $|\vec k|_g$.

We are mainly interested in understanding the qualitative behavior
of the phase diagram of the system  and therefore the approximated values 
of the critical temperature and of the 
critical density can be evaluated  by comparing  the pressure
in the two phases, and determining the transition line at
$p_g  - B = 0$ where
\be
\label{eq3}
p_g = -  T D_g \int \frac{d^3k}{(2\pi)^3} ln[1 - e^{-\beta (E(k) - \mu)}]
\ee
is the gluon pressure, $\mu$ is the gluon ( color independent) chemical potential,
$D_g=16$ is the gluon degeneracy  factor  
and $B$ is the bag constant  that in the pure gauge theory one identifies 
with  $<\alpha_s/\pi G_{\mu \nu}^a G^{a \mu \nu}>$.
At $\mu = 0 $ , the critical temperature turns out to be $ T_c \simeq 0.29$ Gev for $<\alpha_s/\pi G^2>_0 \simeq 0.005$ Gev$^4$,
which is smaller than the average value obtained by QCD sum rules but still within the phenomelogical 
uncertainty \cite{gluoncondensate}.
 
By increasing the chemical potential
one obtains the critical lines in the $T-\mu$ plane, depicted in Fig.1. 
At a certain value of the temperature, $T_c^*$,
the critical line reaches the maximum value of of the chemical potential 
allowed by Eq.(\ref{eq3}),  $\mu_c=E^*_g$. In fact for $\mu > \mu_c$ 
the integral in  Eq.(\ref{eq3}) is ill defined. Then, according to the 
standard picture of the Bose-Einstein condensation (see e.g. \cite{haber}),
for $T< T_c^*$ and $\mu = \mu_c$, which corresponds to the vertical ending piece
of the critical line in Fig.1, gluons progressively condense in the state of 
minimum energy $E^*_g$. When  the temperature reaches the point $T=0$, the totality 
of the gluons is in this state. Unlike the standard condensation where the condensed bosons 
carry no momentum, in this framework the gluons, after condensation, have non vanishing momentum    
$ |\vec k|_g= M$.

\begin{figure}
{{\epsfig{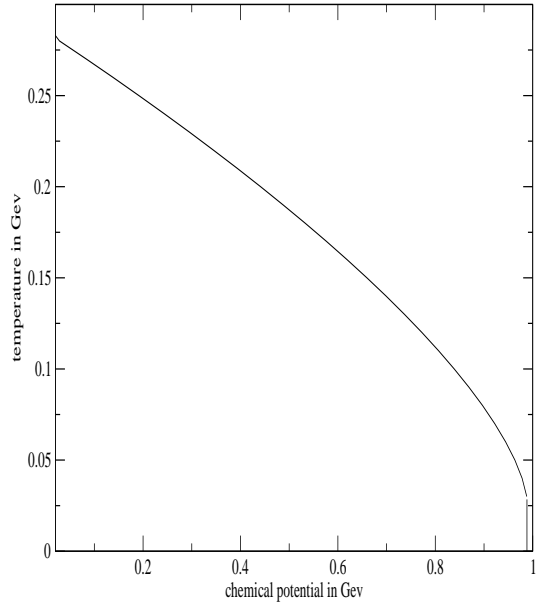}}
\caption{Critical line in the $T-\mu$ plane. The minimum critical temperature corresponds to $\mu=\sqrt{2}M$.  
}}
\end{figure}

The energy density of the system is given by

\be
\epsilon_g =  D_g \int \frac{d^3k}{(2\pi)^3} \frac{1}{[e^{+\beta (E(k) - \mu)}-1]}+B
\ee
and the ``interaction measure'' $(\epsilon_g - 3p_g)/T^4$ can be evaluated and compared with lattice data for $\mu=0$.

The comparison is shown in Fig.2 where our results, the green points,are plotted together with 
those obtained in \cite{zwanziger1} and the lattice findings.
Our points are evaluated for $T > 1.2 T_c$, since a quasi-particle approach is unreliable near the critical point 
and the introduction of the gluon condensate clearly improves the agreement with respect to 
the case $B=0$ \cite{zwanziger1}.

\begin{figure}
{{\epsfig{file=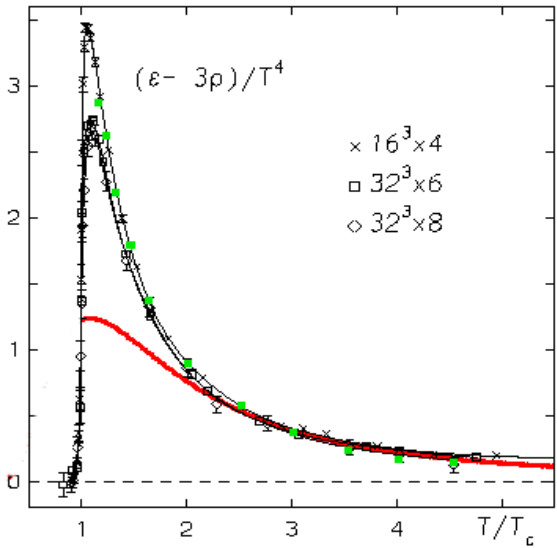,height=8.0 true cm,width=8.0 true cm, angle=0}}
\caption{ Comparison of $(\epsilon_g - 3p_g)/T^4$ evaluated from  Eqs.({\ref{eq1},\ref{eq2})(green points) 
with lattice data, for  $T>1.2 T_c$ .
The red points correspond to the calculations with $B=0$ \cite{zwanziger1}. 
}}}
\end{figure}

The  results in Fig.1 have been obtained by considering  a $\mu$ independent condensate $<\alpha_s/\pi G^2>_0$.
In fact the effect of a finite density on the value of the condensate shoud be taken into account.
Some indications  of the influence of a finite density   can be obtained 
by following the analysis in performed in  \cite{bcz} where $<\alpha_s/\pi G^2>_\mu$ is estimated 
for a system with finite barion number.
As a  rough evaluation  we considered the same density dependence of ref. \cite{bcz} 
including the color-flavor factor
$11/3 =11(N_c N_f)/(11N_c-2N_f)$ with the initial value at $\mu=0$ corresponding to the 
previously used value $B = <\alpha_s/\pi G^2>_0 \simeq 0.005$ Gev$^4$.
There is no qualitative difference with respect to the previous 
case but since $B(\mu)$ decreases by increasing the density, the
transition line is obviously characterized by a smaller critical chemical potential.

In our opinion, the  previous considerations give
 qualitative but clear indications  that there is a critical line 
in the $T - \mu$ plane where  Eq.(\ref{eq1})
leads to a phase transition to a inhomogeneous condensate. 
 As firstly realized by Brazovskii, 
for systems in which the fluctuation spectrum has a minimum at a non zero momentum, $p_c$,
there is a first order transition to a stripe phase, i.e. 
a periodic ordered state with spatial period $2\pi / p_c$.
In Brazovkii's approach the  minimum
in the  inverse propagator at non zero momentum is determined by
a self-consistent Hartree approximation and by expanding 
$E(k)$ around its minimum one obtains an effective lagrangian. 
For example, the Brazovskii-like 
 4-dimensional effective lagrangian
 for a complex scalar field can be written as \cite{gubser,condensed} 
\be
S_{eff}= \int d^4x (\alpha |\partial^2 + p_c^2)\phi|^2 +\beta |\phi|^2 + \frac{\gamma}{2} |\phi|^4)
\ee
that for $\beta < 0$ has a classical minimum at 
$\phi = A$ exp$(ipx)$ with $|A|^2= -\beta/\gamma$ and $|p|=p_c$.
In the present analysis, the minimum in $E(k)$ is due to the QCD infrared 
dynamics \cite{zwanziger1,zwanziger3}.

On the other hand, the use of a  quasiparticle approach has strong limitations. For example, since we discuss
a phase transition to an inhomogeneous phase,  the general expression of the pressure,
$p=T \partial ln Z/ \partial V$, should be used rather than  Eq.(\ref{eq2}) which is valid for homogeneous systems.
In this respect, we are practically working under the assuption of  a slowly varying background
and a definite answer on the existence of a stripe phase for a gluon plasma can be addressed only
by lattice simulations. By the correlators method \cite{iquattro} a  possible signal for this transition 
could be that, at finite density, some gauge invariant QCD
correlator,$C$, shows an oscillating
behaviour $ C= a cos(p_cx)$ corresponding to a macroscopic occupation of the modes $|\vec k|= \pm p_c$.

There is also the question about the
surviving of this phase when  quarks are taken into account. At large quark chemical potential there is the transition to a 
color crystalline phase  with an non uniform colored condensate. Therefore if the gluon condensation occurs at zero 
momentum one should obtain a LOFF fermionic phase with a superimposed gluon uniform phase. This seems  unlikely 
due to quark-gluon  interaction. From this point of view, future  lattice  results  which support the stripe phase for pure gauge 
theory can give an indirect indication of the existence of the color crystalline phase at large quark chemical potential where
lattice calculations are still unreliable.   
 
In conclusion, by considering the non perturbative effects associated with the fundamental modular region, 
we propose a new phase of a Gluon Plasma at finite  density which  corresponds to the 
transition from glueballs to non perturbative gluons  which condense at a non vanishing momentum.
It will be interesting to verify  if also this phenomenon is shared between field theory and condensed matter as
happens for Bose-Eintsein condensation and spontaneous symmetry breaking, superconductivity and chiral symmetry, LOFF phase and
quarks color crystalline phase.

\vskip  10 pt

\noindent
{\bf Acknowledgements} 
The authors thank M. Baldo, G.Nardulli and H.Satz for useful comments.


\begin{thebibliography}{99}

\bibitem{boyd1} G.Boyd et al., Phys. Rev. Lett. {\bf 75}(1995) 4169.

\bibitem{boyd2} G.Boyd et al., Nucl. Phys. {\bf B 469}(1996) 419.

\bibitem{kac} O.Kaczmarek et al., Phys. Rev.{\bf D 70}(2004) 074505.

\bibitem{aoki1} Y.Aoki et al., Phys. Lett. {\bf B 643}(2006) 46.

\bibitem{aoki2} Y.Aoki et al., Nature {\bf 443}(2006) 675.

\bibitem{karsch1} F.Karsch and E.Ledermann,   (2003) hep-lat/0305025.

\bibitem{asakawa1} M.Asakawa, T.Hatsuda and Y. Nakahara,  Nucl. Phys. {\bf A 715}(2003) 863.

\bibitem{asakawa2} M.Asakawa and  T.Hatsuda,  Nucl. Phys. {\bf A 721}(2003) 869.

\bibitem{digiacomo1} M.D'Elia, A.Di Giacomo and E.Meggiolaro,  Phys. Lett.{\bf B 408}(1997) 315.

\bibitem{digiacomo2} M.D'Elia, A.Di Giacomo and E.Meggiolaro,  Phys. Rev. {\bf D 67}(2003)114504.

\bibitem{now} This has been shown by using Nambu - Jona Lasinio model with color charge, see 
\cite{rev,roberto1,raja1,raja2,ciminale,fuku,max1}.

\bibitem{rev} K.Rajagopal and F.Wilczek hep-ph/0011333, M.G.Alford Ann. Rev. Nucl. Part. Sci. {\bf 51}(2001)131,
G.Nardulli Riv. Nuovo Cimento {\bf 25 N3}(2002)1. For more recent reviews: T.Schafer hep-ph/0509068;
M.Alford and K.Rajagopal hep-ph/0606157.

\bibitem{roberto1} R.Casalbuoni et al.,  Phys. Lett. {\bf B 627}(2005) 89.

\bibitem{raja1} K.Rajagopal and R.Sharma, Phys. Rev.{\bf D 74}(2006) 094019.

\bibitem{raja2} K.Rajagopal and R.Sharma, J. Phys. G {\bf  32}(2006) S483.

\bibitem{ciminale} M.Ciminale et al., Phys. Lett. {\bf B 636}(2006) 317.

\bibitem{fuku} K.Fukushima, Phys. Rev.{\bf D 73}(2006) 094016.

\bibitem{max1} M.Mannarelli, K.Rajagopal and R.Sharma, hep-ph/0702021.

\bibitem{levai} P.Levai and U.W.Heinz, Phys. Rev.{\bf C 57}(1998) 1879.

\bibitem{pescier} A.Peshier and W. Cassing, Phys. Rev. Lett. {\bf 94}(2005) 172301.

\bibitem{paolo1} P.Castorina and M.Mannarelli, Phys. Lett. {\bf B 336}(2007) 336.

\bibitem{paolo2} P.Castorina and M.mannarelli, hep-ph/0701206.

\bibitem{zwanziger1} D.Zwanziger, Phys. Rev. Lett. {\bf 94}(2005) 182301.

\bibitem{gribov} V.Gribov,  Nucl. Phys. {\bf B 139}(1978)1.

\bibitem{zwanziger2} D.Zwanziger,  Nucl. Phys. {\bf B 485}(1997)185.

\bibitem{zwanzigern} J.Greensite,S.Olejnik and D.Zwanziger, 
hep-lat/0411032, talk at ``Quark Confinement and the Hadron Spectrum VI'',
 Villasimius, Italy, Sept. 2004.

\bibitem{brazovsky} S.A.Brazovski, Sov. Phys. - JETP    {\bf 41}(1975) 85.

\bibitem{condensed} See for example, P.C. Hohenberg and J.B.Swift,  Phys. Rev.{\bf E 52}(1995) 1828.

\bibitem{verifica} L.Leibler, Macromolecules  {\bf 13}(1980)1602; G.H.Fredrickson and E.Helfand, J.Chem.Phys. 
{\bf 87 }(1987)697; F.S.Bates et al. Phys. Rev. Lett. {\bf 61}(1988) 2229;
 F.S.Bates et al. J.Chem.Phys. {\bf 92 }(1990)6225.

\bibitem{gubser} S.S.Gubser and S.L.Sondhi, Nucl. Phys. {\bf B 605}(2001) 395.

\bibitem{noi} P.Castorina and D.Zappal\`a, Phys. Rev.{\bf D 68}(2003) 065008; P.Castorina,G.Riccobene and D.Zappal\`a,
Phys. Lett. {\bf A 337}(2005) 3463.

\bibitem{bietenolz} J.Ambiorn and S.Catteral  Phys. Lett. {\bf B 636}(2006) 317, W.Bietenholz,F.Hofeinz and J.Nishimura
Nucl. Phys. Proc. Supp {\bf 119}(2003)941, Fortsch. Phys.  {\bf 51}(2003)745.
 
\bibitem{ken} T.H. Hansson, K. Johnson and C. Peterson, Phys. Rev. {\bf D 26} (1982) 2069.

\bibitem{alessandro} A.Drago,M. Gibilisco and C.Ratti, Nucl. Phys. {\bf A 742}(2004)165. 

\bibitem{gluoncondensate} M.A. Shifman, A.I. Vainshtein and V.I.Zakharov  Nucl. Phys. {\bf B 147}(1979)385, 
Nucl. Phys. {\bf B 147}(1979)448; J.Reinders,H,Rubinstein and S.Yazaki , Phys. Rep. {\bf B 127}(1985). 

\bibitem{haber} H.E. Haber and H.A. Weldon, Phys. Rev. Lett. {\bf 46} (1981) 1497; Phys. Rev.{\bf D 25}(1982) 502.

\bibitem{bcz} M.Baldo, P.Castorina and D.Zappal\`a,  Nucl. Phys. {\bf A 743}(2004)3.

\bibitem{zwanziger3} D.Zwanziger,  hep-ph/0610021.

\bibitem{iquattro} A. Di Giacomo, H.G. Dosch,V.I.Shevchenko amd Y.A. Simonov, Phys. Rep {\bf 372}(2002)319.

\end{thebibliography}
\end{document}